\documentclass{ws-procs961x669}            

\usepackage{hyperref}
\usepackage{amsfonts}
\usepackage{mathrsfs}
\usepackage{amssymb}
\usepackage{graphicx, epsfig, amsmath}
\usepackage{dcolumn}
\usepackage{bm}

\newcommand{\la}{\langle}
\newcommand{\ra}{\rangle}
\newcommand{\w}{\omega}

\newcommand{\be}{\begin{equation}}
\newcommand{\ee}{\end{equation}}
\newcommand{\bea}{\begin{eqnarray}}
\newcommand{\eea}{\end{eqnarray}}
\newcommand{\bes}{\begin{subequations}}
\newcommand{\ees}{\end{subequations}}

\begin{document}
\title{Stress-energy Tensor for a Quantized Scalar Field in a Four-Dimensional Black Hole Spacetime that Forms From the Collapse of a Null Shell}

\author{Shohreh Gholizadeh Siahmazg$^*$,  Paul R. Anderson $^{**}$, and Raymond D. Clark}

\address{Department of Physics, Wake Forest University,\\
Winston Salem, North Carolina 27109, USA\\
$^*$E-mail: ghols18@wfu.edu\\
$^{**}$ E-mail: anderson@wfu.edu
}

\author{Alessandro Fabbri}

\address{Departamento de F\'isica Te\'orica and IFIC, Universidad de Valencia-CSIC\\
C. Dr. Moliner 50, 46100 Burjassot\\
E-mail: afabbri@ific.uv.es}

\begin{abstract}
A method is presented which allows for the numerical computation of the stress-energy tensor for a quantized massless minimally coupled scalar field in the region outside the event horizon of a 4D Schwarzschild black hole that forms from the collapse of a null shell. This method involves taking the difference between the stress-energy tensor for the {\it in} state in the collapsing null shell spacetime and that for the Unruh state in Schwarzschild spacetime. The construction of the modes for the {\it in} vacuum state and the Unruh state is discussed. Applying the method, the renormalized stress-energy tensor in the 2D case has been computed numerically and shown to be in agreement with the known analytic solution. In 4D, the presence of an effective potential in the mode equation causes scattering effects that make the construction of the in modes more complicated. The numerical computation of the {\it in} modes in this case is given.
\end{abstract}


\bodymatter

\section{Introduction}\label{aba:sec1}
The expectation value of the renormalized stress-energy tensor operator for a quantized field is a useful way to study quantum effects curved space. It can also be used in the context of semiclassical gravity to compute the backreaction of the quantum field on the background geometry. In the case of four-dimensional, 4D, black holes, this quantity  has to date only been computed for static black holes ~\cite{fawcett,howard-candelas,howard,jensen-ottewill,jensen-et-al,ahs1,ahs2,ahl,choag,abf,breen-ottewill,levi-ori,levi,Zilberman-Levi-Ori} and Kerr black holes ~\cite{duffy-ottewill,levi-et-al-kerr}. However, to our knowledge, a full numerical computation of this quantity has not been done for a quantized field in a 4D spacetime in which a black hole forms from the collapse of a null shell, which is probably the simplest model for the formation of a black hole. 

In Ref.~\citenum {ourpaper}, we developed a method to numerically compute the full renormalized stress-energy tensor for a massless minimally coupled scalar field in the case of a spherically symmetric black hole in 4D that forms from the collapse of a null shell. This method can be used in the region outside the null shell and future horizon, where by Birkhoff's theorem, the geometry is described by the Schwarzschild metric. 

In this proceeding, we review this method with a focus on the computation of a complete set of {\it in} modes that can be used to construct the quantum field in the region outside the null shell.  We also present new numerical results for a low frequency {\it in} mode on the future horizon and for a mode with relatively high frequency on the part of the future horizon close to the null shell trajectory.

In Sec. 2, we review the null shell spacetime and the metrics describing the geometry inside and outside of the null shell. In Sec. 3, we discuss the quantization of the massless minimally coupled scalar field in the null shell spacetime. In Sec. 4, we present our method to expand the {\it in} modes in terms of a complete set of modes in pure Schwarzschild spacetime and present our numerical results for the high and low frequency modes on the future horizon.  In Sec. 5, a proper method  for the renormalization of the stress-energy tensor is given. In this section, we summarize the application of our method in Ref.~\citenum{ourpaper} to the case of a collapsing null shell spacetime which has a perfectly reflecting mirror at the spatial origin. 

\section{Collapsing null shell}
\begin{figure}[h]
\begin{center}
 \parbox{2.1in}{\includegraphics[trim={5cm 19cm 5cm 2cm},clip, width=3in]{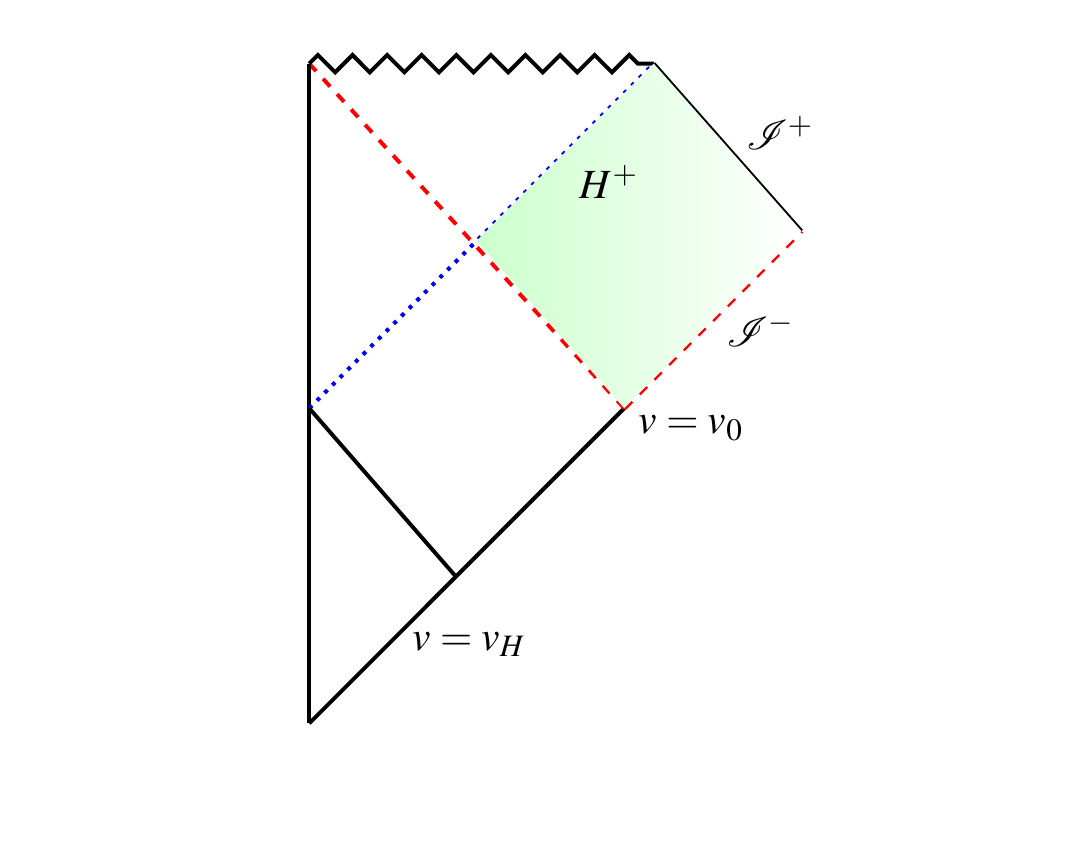}}
\end{center}
\caption{Penrose diagram for a spacetime in which a null shell collapses to form a spherically symmetric black hole. The vertical line on the left corresponds to the surface $r=0$ which is also the surface where $u=v$.  The dashed red
line on $v=v_0$ is the trajectory of the null shell. The horizon, $H^+$ is the dotted blue curve. Inside the shell trajectory $H^+$ corresponds to the surface $u=v_H$ and outside the shell trajectory it corresponds to $u_s=\infty$. A Cauchy surface is shown by the dashed line. It is the union of the surface $v=v_0$ with the part of $\mathscr{I^{-1}}$ with $v>v_0$.}
\label{sg:fig1}
\end{figure}

The model we consider is a spherically symmetric null shell whose collapse results in the formation of a black hole. The Penrose diagram of the spacetime is depicted in Fig. \ref{sg:fig1} The spacetime inside the null shell is described by the flat metric
\[
ds^2 = -dt^2 + dr^2 + r^2 d \Omega^2    \;,
\]
and from Birkhoff's theorem, the metric outside the shell is the Schwarzschild metric
\[
  ds^2 = -\left(1-\frac{2M}{r} \right) dt_s^2
+ \left( 1-\frac{2M}{r} \right)^{-1} dr^2 + r^2 d \Omega^2  \;
\]
with $d\Omega^2=d\theta^2+\sin^2{\theta}d\phi^2$.
It is more convenient to use radial null coordinates to match the geometries inside and outside of the shell. In the interior,
 \[u=t-r\;, \quad v=t+r\;.
 \]
and in the exterior region,
\[u_s=t_s-r_*\;, \quad v=t_s+r_*\;,
 \]
where  $ r_*=r+2M\ln{\big(\frac{r-2M}{2M}}\big)$ is the tortoise coordinate in Schwarzschild spacetime. The null shell trajectory is $v=v_0$. We match the two spacetimes so that the $v$ coordinate and the angular coordinates are continuous across the null shell trajectory. Applying this condition gives the following relation between the $u$ an $u_s$ coordinates  \cite{Fabbri:2005mw, m-p}
 \[u_s=u-4M\log(\frac{v_H-u}{4M}),
 \]
where $v_H=v_0 -4M$.
\section{Massless minimally coupled scalar field}
We consider a massless minimally coupled scalar field in the null shell spacetime.   In a general  static spherically  symmetric  spacetime, it can  be  expanded in the following way,
\[\phi = \sum_{\ell = 0}^\infty  \sum_{m = -\ell}^\ell \int_0^\infty [a_{\omega \ell m} f_{\omega \ell m} + a^\dagger_{\omega \ell m} f^{*}_{\omega \ell m}] \quad 
\] 
with $\Box\; f_{\omega \ell m} = 0$. In the region inside the null shell, $v<v_0$, separation of variables gives
\begin{equation}
f_{\omega \ell m}=\frac{Y_{\ell m}(\theta,\phi)}{r \sqrt{4 \pi \omega}} \psi_{\omega \ell}(t,r)=\frac{Y_{\ell m}(\theta,\phi)}{r \sqrt{4 \pi \omega}} e^{-i \omega t} \chi_{\omega \ell}(r),\end{equation}
while in the region outside the null shell, $v>v_0$, it gives 
\begin{equation}f_{\omega \ell m}=\frac{Y_{\ell m}(\theta,\phi)}{r \sqrt{4 \pi \omega}} \psi_{\omega \ell}(t_s,r)=\frac{Y_{\ell m}(\theta,\phi)}{r \sqrt{4 \pi \omega}} e^{-i \omega t_s} \chi_{\omega \ell}(r).\end{equation}
In the regions $v<v_0$  and $v>v_0$ respectively, the radial parts of the mode functions satisfy the differential equations
\begin{align}\frac{d^2 \chi_{\omega \ell}}{d r^2}& = - \left[\omega^2 - \frac{\ell(\ell+1)}{r^2} \right] \chi_{\omega \ell}.\\
\frac{d^2 \chi_{\omega \ell}}{d r_{*}^2}& = -  \left[\omega^2 - \left(1 - \frac{2M}{r} \right) \left( \frac{2 M}{r^3}+  \frac{\ell(\ell+1)}{r^2} \right) \right] \chi_{\omega \ell}\;.
\end{align}
The {\it in} state is fixed by requiring that $\psi_{\w \ell}=e^{-i\w v}$ on past null infinity and it vanishes at $r=0$. The solution with these properties has the form
\begin{equation}\psi_{\omega \ell}^{in}(r,t)=C_{\ell} e^{i \w t}\omega r j_{\ell}(\w r)\label{sg:eq5}
\end{equation}
inside the null shell, where $C_{\ell}$ is fixed by the aforementioned condition on past null infinity.
Here $j_{\ell}$ is a spherical bessel function of the first kind. It is not possible for this solution to have the form  $e^{-i\w t_s}\chi_{\w \ell}(r)$ outside the null shell. The solution in this region is more complicated.
\section{Computation of $f^{in}_{\omega \ell m}$}
\begin{figure}[h]
\begin{center}
 \parbox{2.1in}{\includegraphics[width=2.5in]{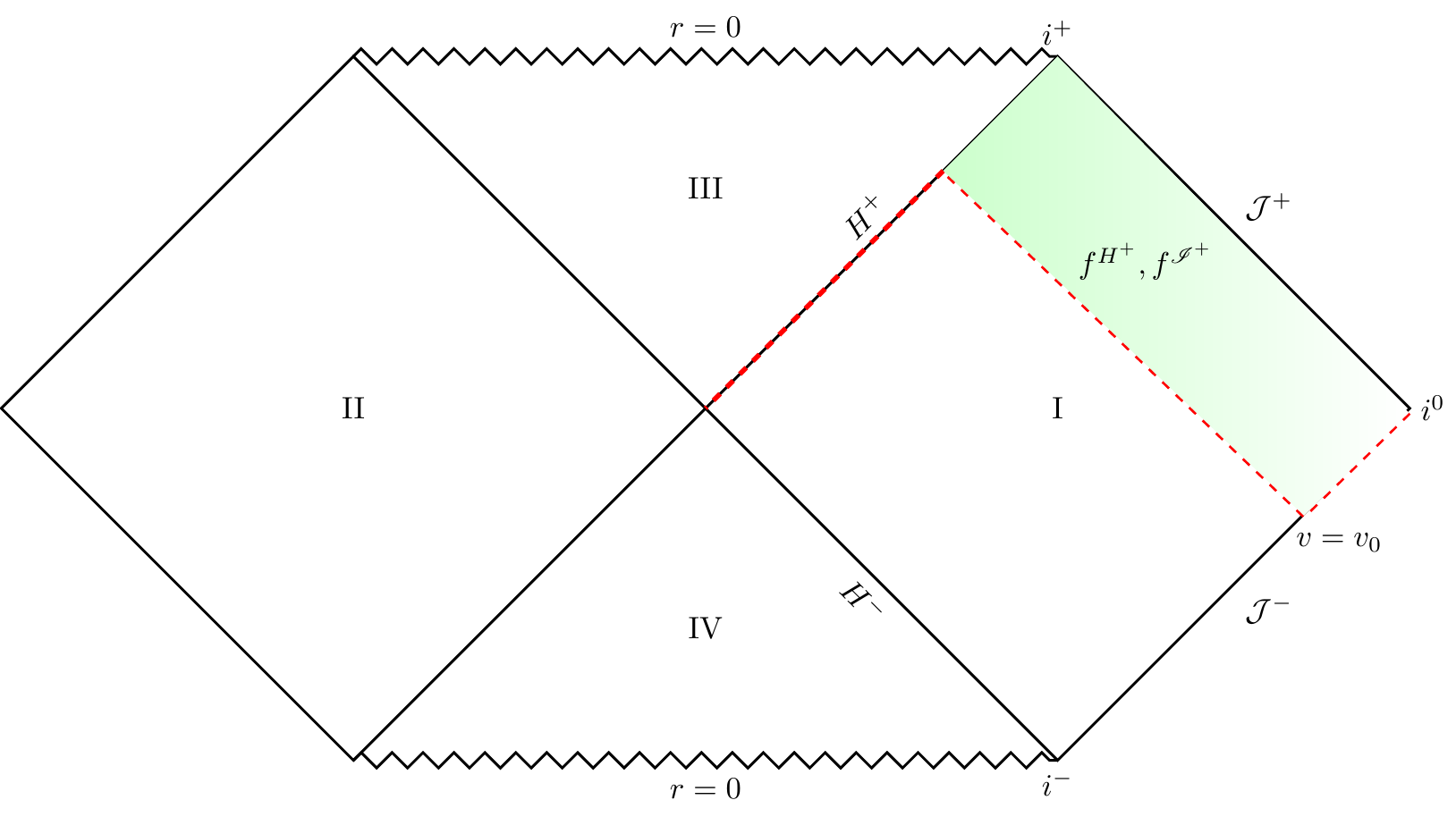}}
\end{center}
\caption{Penrose diagram for Schwarzschild spacetime showing the Cauchy surface used for matching the in modes in the null shell spacetime to a complete set of modes in Schwarzschild spacetime in the region outside the past and future horizons. The Cauchy surface is denoted by the dashed red curve}
\label{sg:fig2}
\end{figure}

We can compute $f^{in}_{\w \ell m}$ outside the null shell and the event horizon by expanding it in terms of a complete set of modes since the geometry here is the Schwarzschild geometry. This problem can be mapped to the shaded part of pure Schwarzschild spacetime shown in Fig. \ref{sg:fig2}.
 We choose a complete sets of modes that consists of the union of the modes $f^{H^+}_{\w \ell m}$ that are positive
frequency on future horizon and zero on future null infinity, and the modes  $f^{\mathscr{I}^+}_{\w \ell m}$ that are positive frequency on the future null infinity and zero on the future horizon, $i.e$.,
\begin{align}
f^{in}_{\w \ell m}& =\sum_{\ell'=0}^{\infty}\sum_{m'=-\ell'}^{\ell'} \int_{0}^{\infty} d \w' \Big[A^{\mathscr{I}^{+}}_{\w \ell m \w'\ell' m'}\;f^{\mathscr{I}^+}_{\w' \ell' m'}+B^{\mathscr{I}^{+}}_{\w \ell m \w'\ell' m'}\;(f^{\mathscr{I}^+}_{\w' \ell' m'})^*\notag\\&+A^{H^{+}}_{\w \ell m \w'\ell' m'}\;f^{H^+}_{\w' \ell' m'}  + B^{H^{+}}_{\w \ell m \w'\ell' m'}\;(f^{H^+}_{\w' \ell' m'})^*\Big].\label{sg:eq6}
\end{align}
The matching coefficients $A^{\mathscr{I}^{+}}_{\w \ell m \w'\ell' m'}$, $B^{\mathscr{I}^{+}}_{\w \ell m \w'\ell' m'}$, $A^{H^{+}}_{\w \ell m \w'\ell' m'}$, and $B^{H^{+}}_{\w \ell m \w'\ell' m'}$ can be found using the following scalar products on the Cauchy surface shown in Fig. \ref{sg:fig2}.
\begin{align}
A^{(\mathscr{I^+}, H^+)}_{\w \ell m\w' \ell' m'} &=( f^{\rm in}_{\w \ell m} , f^{(\mathscr{I}^+, H^+)}_{\w' \ell' m'}) \label{sg:eq7}, \\
B^{(\mathscr{I^+}, H^+)}_{\w \ell m\w' \ell' m'} &= -( f^{\rm in}_{\w \ell m} , (f^{(\mathscr{I}^+, H^+)}_{\w' \ell' m'})^{*}).\label{sg:eq8}
\end{align}
The reason one can expand the $in$ modes in this way is that the same  differential equations govern the  evolution of the $in$ modes in the shaded region in Fig. \ref{sg:fig1} and the shaded region in Fig. \ref{sg:fig2} because the metric is the same in both regions. However, one may notice that the union of the part of past null infinity with $v>v_0$ and the null shell surface does not form a Cauchy surface in pure Schwarzschild spacetime. We resolve this issue by adding the part of the future horizon with $-\infty<v<v_0$ to  the union of $\mathscr{I^-}$ with $v>v_0$ and the null shell surface $v=v_0$, as shown in Fig. \ref{sg:fig2}.  We also need to specifity $\psi_{\w \ell}^{in}$ on the Cauchy surface to evaluate the scalar products in Eqs. \ref{sg:eq7} and \ref{sg:eq8}. On the part of the Cauchy surface with $v>v_0$ on past null infinity, $\psi^{in}_{\w \ell}=e^{-i\w v}$ and on the part where $v=v_0$ , $\psi^{in}_{\w \ell}$ is given by Eq. \ref{sg:eq5}. For the part with $v<v_0$ on the future horizon, we can specify $\psi^{in}_{\w \ell}$ any way we like so long as it is continuous at $v=v_0$.

Before computing the matching coefficients, we introduce a different complete set of modes that are defined by two linearly independent solutions to the radial mode equation in Schwarzschild spacetime with the following properties
\begin{align}
\chi^\infty_R &\to e^{i \w r_*} \;,  \qquad r_* \to \infty, \\
\chi^\infty_L &\to e^{ -i \w r_*} \;, \qquad r_* \to \infty.
\end{align}
Near the horizon, they have the behaviors  ~\cite{rigorous}
\begin{align}
 \chi^\infty_R &\to E_R(\w) e^{i \w r_*} + F_R(\w) e^{-i \w r_*} \;, \qquad r_* \to -\infty, \\
      \chi^\infty_L &\to E_L(\w) e^{i \w r_*} + F_L(\w) e^{-i \w r_*} \;, \qquad r_* \to - \infty.
\end{align}
where $E_R$, $E_L$, $F_R$, and $F_L$ are scattering parameters that can be determined numerically.

\begin{figure}[h]
\begin{center}
 \parbox{2.1in}{\includegraphics[width=2.5in]{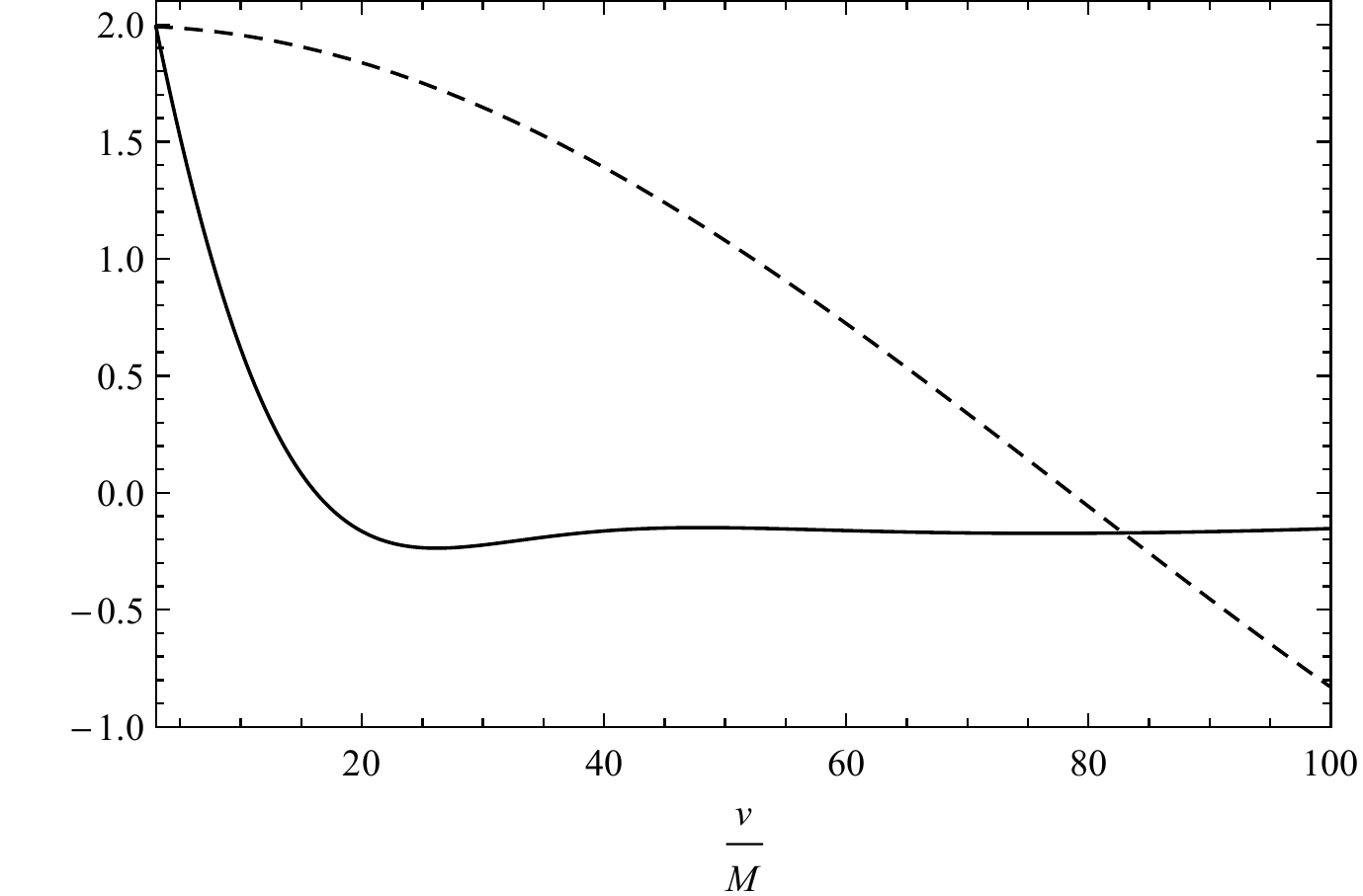}}
 \hspace*{16pt}
 \parbox{2.1in}{\includegraphics[width=2.5in]{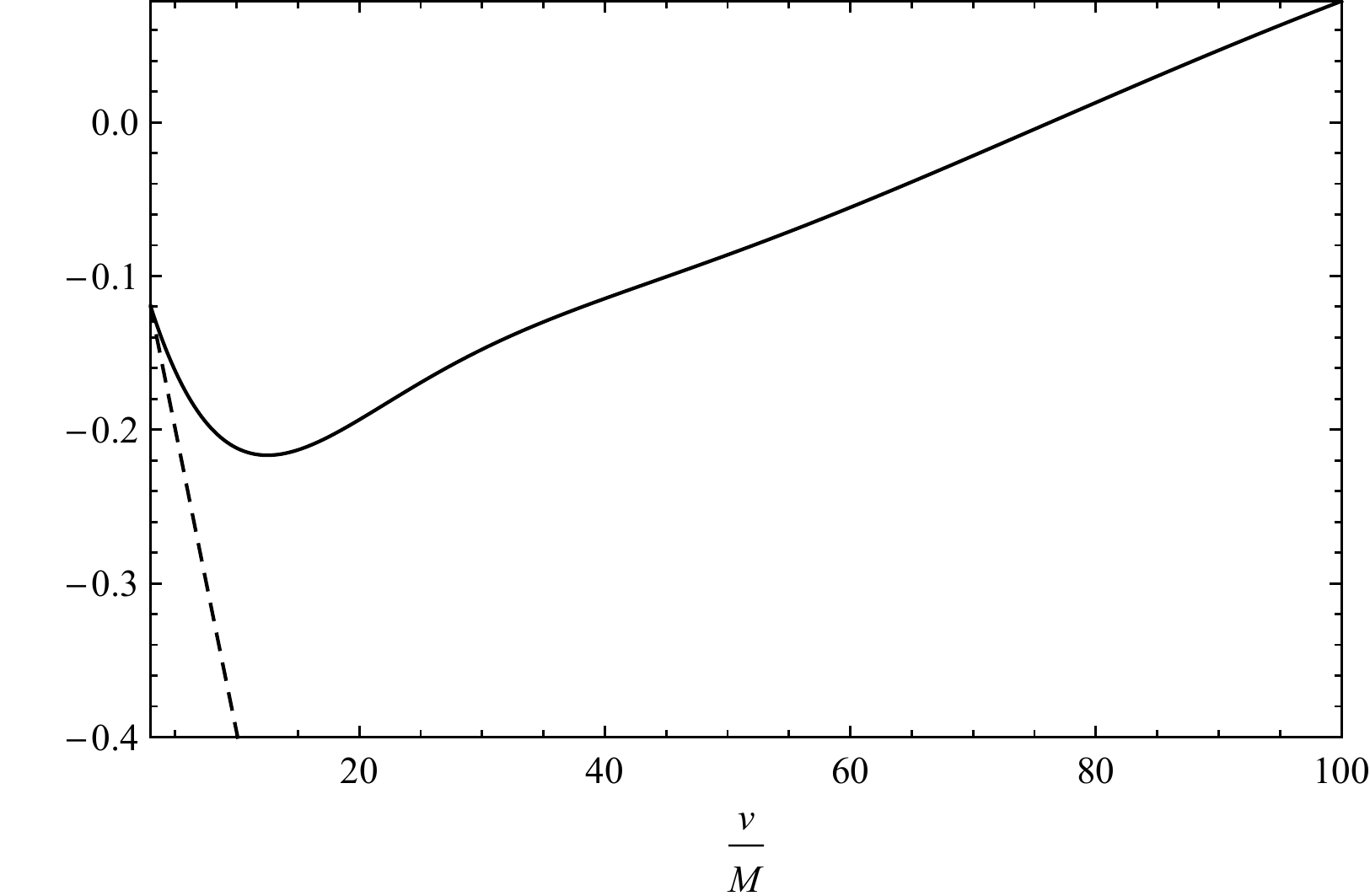}}
\end{center}
\caption{Real part (left) and imaginary part (right) of $\psi^{in}_{\w 00}(v)$ on $H^+$ for $v>v_0$. $\frac{v_0}{M}=3$ and $M\w=0.02$. The dashed lines and solid lines correspond to the $in$ modes in the 2D and the 4D cases respectively.}
\label{sg:fig3}
\end{figure}

\begin{figure}[h]
\begin{center}
 \parbox{2.1in}{\includegraphics[width=2.5in]{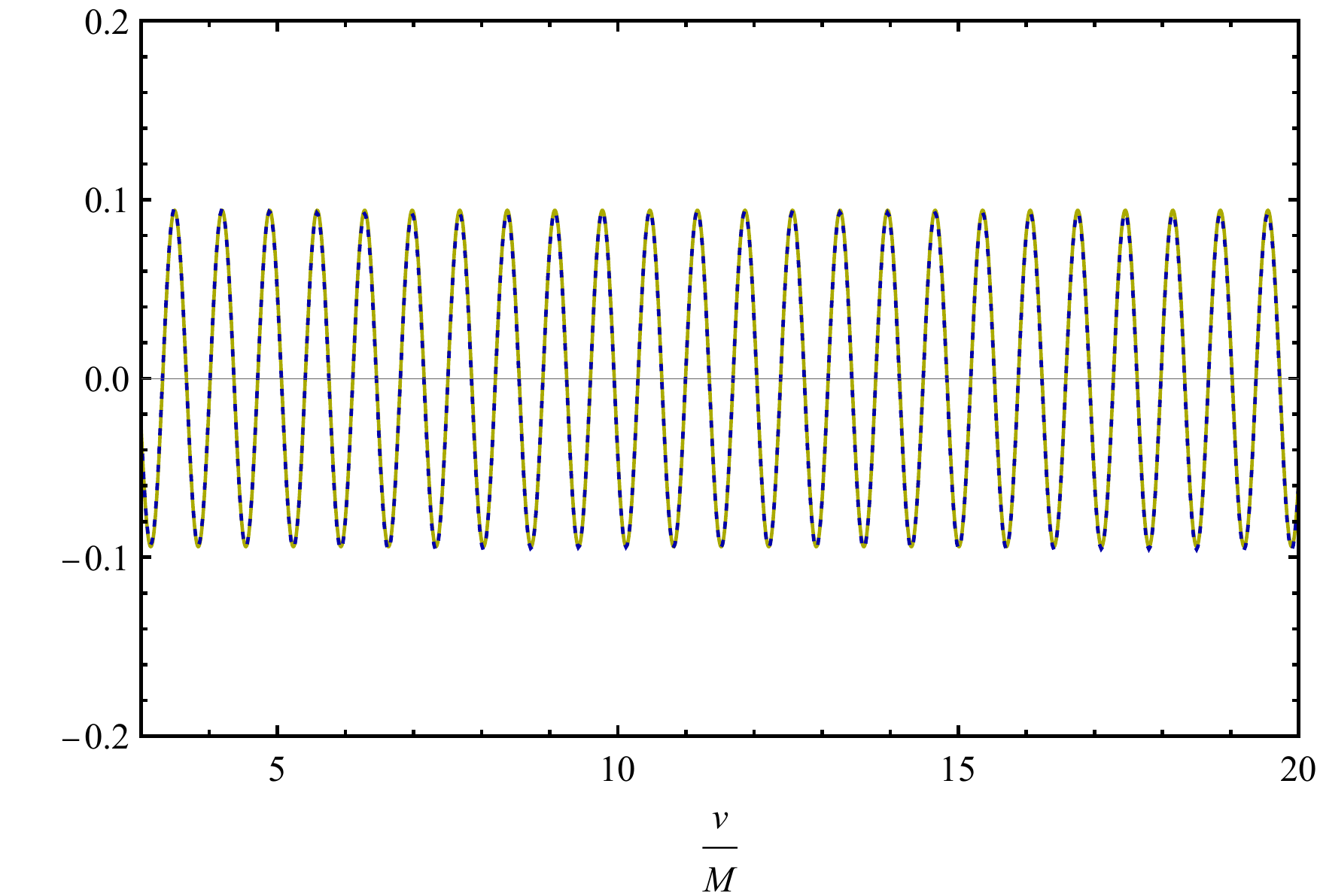}}
 \hspace*{16pt}
 \parbox{2.1in}{\includegraphics[width=2.5in]{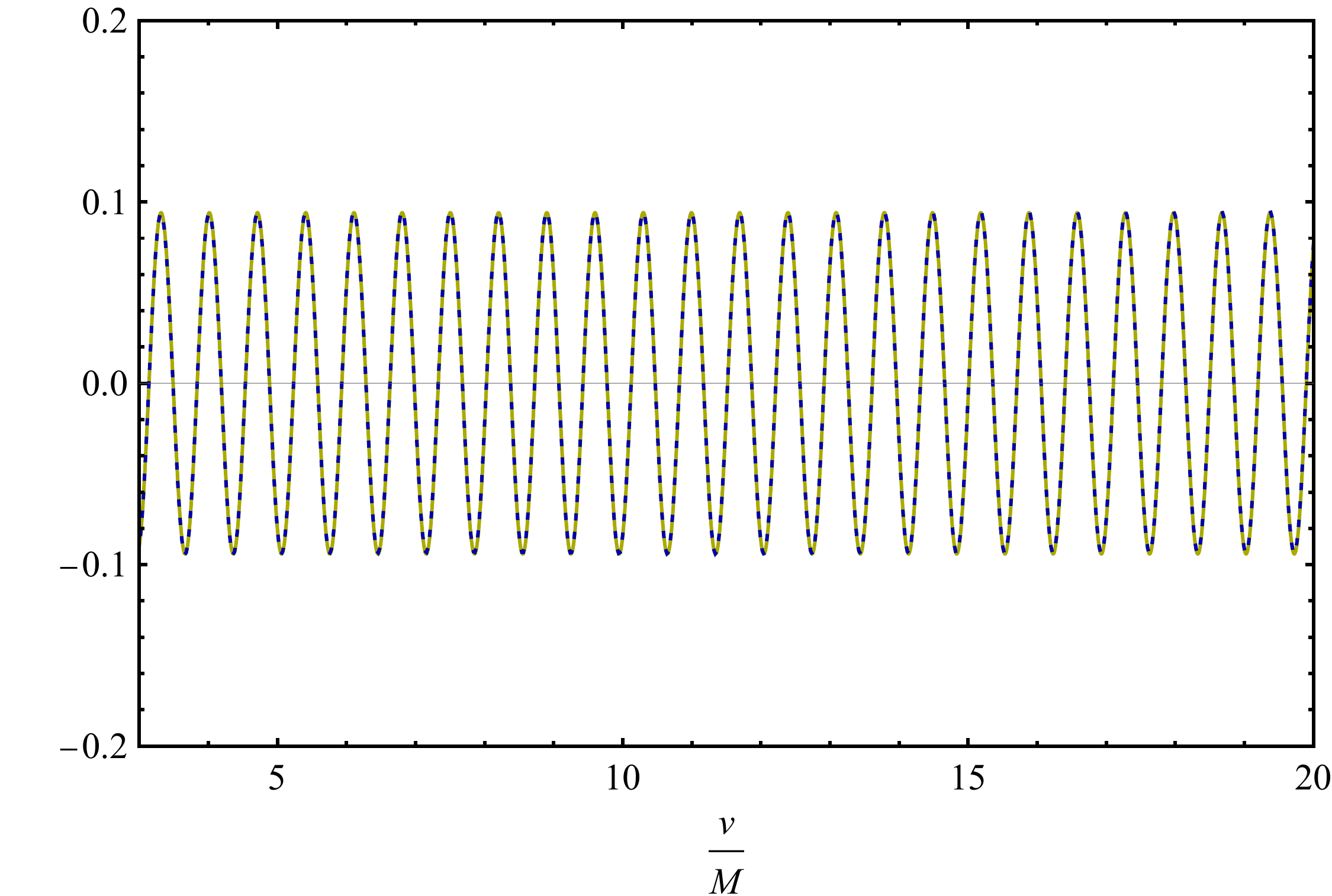}}
\end{center}
\caption{Real part (left) and imaginary part (right) of $\psi^{in}_{\w 00}(v)$ on $H^+$ for $v>v_0$. $\frac{v_0}{M}=3$ and $M\w=9$. The dashed blue lines and solid yellow lines correspond to the $in$ modes in the 2D and the 4D cases respectively.}
\label{sg:fig4}
\end{figure}

Evaluating the scalar products in Eqs. \ref{sg:eq7} and \ref{sg:eq8} gives the following results for the matching coefficients ~\cite{ourpaper}
\begin{align}
A^{ H^{+}}_{\w \w' l} &= - \frac{i}{2 \pi} \sqrt{\frac{\w'}{\w}} \frac{e^{i \w' v_0}}{\w' - i \epsilon} \psi^{\rm in}_{\w \ell}(v_H,v_0)
   + \frac{i}{2 \pi} \sqrt{\frac{\w'}{\w}} \frac{1}{F_L^{*}(\w',\ell)}\frac{e^{i(\w'-\w)v_0}}{\w'-\w+i \epsilon} \notag \\&
   + \frac{i}{2 \pi \sqrt{\w \w'} } \int_{-\infty}^{v_H} du \, \left[\partial_u \psi^{\rm in}_{\w \ell}(u,v_0) \right] \psi^{H^{+} *}_{\w' \ell}(u_s(u),v_0)\label{sg:eq13}, \\
      B^{ H^{+}}_{\w \w' l} &= \frac{i}{2 \pi} \sqrt{\frac{\w'}{\w}} \frac{e^{-i \w' v_0}}{\w' + i \epsilon} \psi^{\rm in}_{\w \ell}(v_H,v_0)
      - \frac{i}{2 \pi} \sqrt{\frac{\w'}{\w}} \frac{1}{F_L(\w',\ell)}\frac{e^{-i(\w+\w')v_0}}{\w'+\w-i \epsilon} \notag \\ &
       - \frac{i}{2 \pi \sqrt{\w \w'} } \int_{-\infty}^{v_H} du \, \left[\partial_u \psi^{\rm in}_{\w \ell}(u,v_0) \right] \psi^{H^{+}}_{\w' \ell} (u_s(u),v_0)\label{sg:eq14}.
\end{align}
\begin{align}
    A^{ \mathscr{I}^{+}}_{\w \w' l} &=  - \frac{i}{2 \pi} \sqrt{\frac{\w'}{\w}} \frac{F_R^{*}(\w',\ell)}{F_L^{*}(\w',\ell)}\frac{e^{-i(\w-\w')v_0}}{\w'-\w+i \epsilon} \notag \\  & 
       -\frac{i}{2 \pi \sqrt{\w \w'}} \int_{-\infty}^{v_H} du \, \left[  \psi^{\rm in}_{\w \ell}(u, v_0) - e^{-i \w v_0} \right] \partial_u \psi^{\mathscr{I}^+  *}_{\w' \ell}(u_s(u),v_0)\label{sg:eq15},   \\
     B^{ \mathscr{I}^{+}}_{\w \w' l} &=  \frac{i}{2 \pi} \sqrt{\frac{\w'}{\w}} \frac{F_R(\w',\ell)}{F_L(\w',\ell)}\frac{e^{-i(\w+\w')v_0}}{\w'+\w-i \epsilon} \notag \\
  &+ \frac{i}{2 \pi \sqrt{\w \w'}} \int_{-\infty}^{v_H} du \, \left[ \psi^{\rm in}_{\w \ell}(u, v_0) - e^{-i \w v_0} \right] \partial_u \psi^{\mathscr{I}^+}_{\w' \ell}(u_s(u),v_0)\label{sg:eq16}.
\end{align}
In the case $\ell=0$, the {\it in} mode functions have the form $f^{in}_{\w00}=\frac{\psi^{in}_{\w 00}}{r\sqrt{4\pi \w}}$ where $\psi^{in}_{\w 0}=e^{-i\w v}-e^{-i\w u}$ for $v\leq v_0$. We used the {\it v-}dependent terms in the matching coefficients to construct $f^{in}_{\w 00}$ on $H^+$. The numerical results are shown in Fig. \ref{sg:fig3} and Fig. \ref{sg:fig4}. In Fig. \ref{sg:fig3}, the real and imaginary parts of the {\it v-}dependent part of the {\it in} mode function have been numerically computed on $H^+$. The results show that {\it in} mode function is continuous across the null shell as expected.
For large values of $\w$, the effective potential in the mode equation is always small compared to $\w^2$ and one can ignore the scattering effects. Hence, one should expect to see the same behaviour as in the 2D case where there are no scattering effects. This is shown to be correct in Fig. \ref{sg:fig4}. where the real and imaginary parts of $f^{in}_{\w 00}$ are plotted for $M\w=9$.
\section{Stress-energy tensor and renormalization}
One can renormalize the stress-energy tensor by subtracting from the unrenormalized expression for the stress-energy tensor for the in vacuum state, the unrenormalized stress-energy tensor for the Unruh state. Since the renormalization counterterms are local and thus do not depend on the state of the quantum field, this quantity will be finite. Then one can add back the unrenormalized stress-energy tensor for the Unruh state and subtract from it the renormalization counter terms. Schematically one can write
\begin{align}
\la {\rm in}| T_{ab}| {\rm in} \ra_{\rm ren} &= \Delta \la T_{ab} \ra +   \la U| T_{ab} | U \ra_{\rm ren},
\end{align}
where $\Delta \la T_{ab} \ra = \la {\rm in}| T_{ab} |{\rm in} \ra_{\rm unren} - \la U| T_{ab} | U \ra_{\rm unren}$. Note that the Unruh state is defined by a set of modes that are positive frequency on the past horizon with respect to the Kruskal time coordinate and the set of modes  that have the form $\psi_{\w \ell}=e^{-i\w v}$ on past null infinity. The quantity $\la U| T_{ab} | U \ra_{\rm ren}$ has been numerically computed for a massless minimally coupled scalar field in Schwarzschild spacetime. Thus what remains is to compute the difference between the unrenormalized expressions. 

The unrenormalized stress-energy tensor can be computed by taking derivatives of the Hadamard Green's function as follows ~\cite{christensen-76},
\begin{equation}
\la T_{ab} \ra_{\text{unren}} = \frac{1}{4} \lim_{x' \to x} \left[ \left( g_a^{c'} G^{(1)}_{;c'; b}(x,x')+ g_b^{c'}  G^{(1)}_{;a; c'} (x,x') \right)  - \, g_{a b}\, g^{c d'} G^{(1)}_{;c ;d'}(x,x') \right],
\end{equation}
 where the quantity $g_a^{b'}$ parallel transports a vector from $x'$ to $x$ and is called the bivector of parallel transport.
\subsection{Stress-energy tensor in 2D}
\begin{figure}[h]
\begin{center}
 \parbox{2.1in}{\includegraphics[trim={2.5cm 0cm 0cm 3.5cm},clip,width=3.5in]{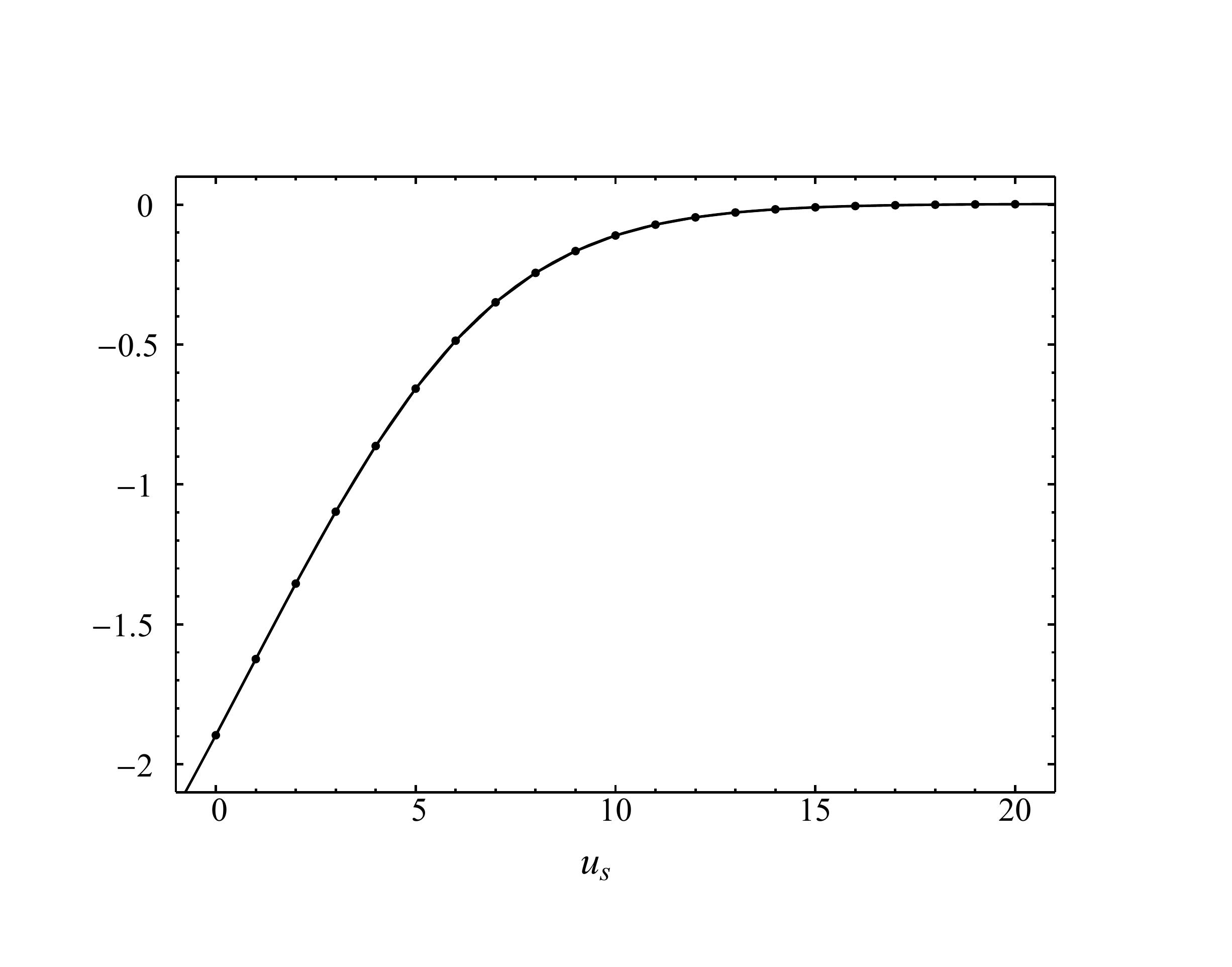}}
\end{center}
\caption{The quantity $10^4 M^2  \langle T_{tt} \rangle$is plotted for the massless minimally coupled scalar field in the region exterior to the null shell and to the event horizon. The dots correspond to the results of the numerical computations. The solid curve represents the analytic results. }
\label{sg:fig5}
\end{figure}

In this section, we show how our method can be applied to the case of a 2D null shell spacetime which has a perfectly reflecting mirror at $r=0$. There is no scattering that means $E_R=F_L=1$ and $E_L=F_R=0$. The matching coefficients are ~\cite{ourpaper},
\begin{align}
A^{\mathscr{I}^{+}}_{\w \w'} &= -\frac{1}{2 \pi} \sqrt{\frac{\w}{ \w'}} (4M)^{i 4M \w'} e^{-i(\w-\w') v_H} \frac{\Gamma(1-i4M\w')}{[-i(\w-\w')+\epsilon]^{1-i4M\w'}}\label{sg:eq19},\\
B^{\mathscr{I}^{+}}_{\w \w'} &=  \frac{1}{2 \pi} \sqrt{\frac{\w}{ \w'}} (4M)^{-i 4M \w'} e^{-i(\w+\w') v_H} \frac{\Gamma(1+i4M\w')}{[-i(\w+\w')+\epsilon]^{1+i4M\w'}}\label{sg:eq20}.
\end{align}
The expression for $f^{in}_{\w}$ can be obtained by substituting Eqs. \ref{sg:eq19} and \ref{sg:eq20} into Eq. \ref{sg:eq6}. Those for  $f^{\text{Unruh}}_{\w}$ can be obtained using similar expressions. See Ref.~\citenum{ourpaper} for more details.
Next, we construct the Hadamard form of the Green's function which in 2D is
\begin{equation}
 G_{in}^{(1)}(x,x') = \int_0^\infty d \w \; [f^{in}_\w(x) f^{in \;*}_{\w}(x') + f^{in \;*}_\w(x) f^{in}_{\w}(x') ]\;.
 \end{equation}
We subtract off the corresponding expression for the Unruh state to obtain
\[\Delta G(x,x')=G^{(1)}_{in}(x,x')-G^{(1)}_{\text{Unruh}}(x,x')\].
which  gives
\begin{align}
\Delta \langle T_{tt} \rangle&=  -(1-\frac{2M}{r}) \lim_{x' \to x} \frac{1}{4}(\Delta G_{;t';r}+\Delta G_{;t;r'}).
\end{align}
Our method results in a complicated operation for $\Delta \langle T_{tt} \rangle$ which initially contains a triple integral. One of the integrals can be computed in closed form with the result ~\cite{ourpaper}
\begin{align}
\Delta \langle T_{tt} \rangle&=\mathfrak{R}\Bigg\{ \frac{i}{8\pi^3}\int_0^{\infty}d\w_1 \; \int_0^{\infty}d\w_2 \; e^{-2\pi M(\w_1+\w_2)}\notag \\ & 
\times \bigg\{ e^{i(\w_2-\w_1) u_s} \frac{(4M\w_1e^{\frac{v_H}{4M}})^{4iM\w_1}}{(4M\w_2e^{\frac{v_H}{4M}})^{4iM\w_2}}\frac{\Gamma(1-4iM\w_1)\Gamma(1+4iM\w_2)}{4M(\w_1-\w_2-i\epsilon)} \notag \\  &
+ e^{-i(\w_2 +\w_1)u_s}(4M\w_1e^{\frac{v_H}{4M}})^{4iM\w_1}(4M\w_2e^{\frac{v_H}{4M}})^{4iM\w_2} \notag \\ & 
 \times \frac{\Gamma(1-4iM\w_1)\Gamma(1-4iM\w_2)}{4M(\w_1+\w_2)}\bigg\} \Bigg\}\;.
\end{align}
This quantity has been computed numerically \cite{ourpaper} and the results are shown in Fig. \ref{sg:fig5}. The stress-energy tensor for a massless minimally coupled scalar field in the 2D collapsing null shell spacetime has been previously computed analytically using a different method ~\cite{Fabbri:2005mw, m-p,mirror-bh,late-time, hiscock} and the stress-energy tensor for the Unruh state has also been computed analytically ~\cite{Fabbri:2005mw, m-p,mirror-bh,late-time, hiscock, Hawking:1974sw, Unruh-1976, elster, akhmedov-godazgar-popov, Davies-Fulling-Unruh} . Our results are shown with the dots in Fig. \ref{sg:fig5} and the result found by using previous methods is shown with a solid curve . They agree to more than ten digits.
 It is worth mentioning that in 2D, once $
\Delta \langle T_{tt} \rangle$is numerically computed, $\Delta \langle T_{rr} \rangle$ and $\Delta \langle T_{rt} \rangle$ can be easily determined ~\cite{ourpaper}.
\section{Summary}
We have reviewed a method of computing the stress-energy tensor for a massless minimally-coupled scalar field in a spacetime in which a spherically symmetric black hole is formed by the collapse of a null shell. This method primarily involves two parts. One part is the expansion of the {\it in} mode functions in terms of a complete set of modes in the part of a Schwarzschild black hole that is outside the event horizon. The matching coefficients of the expansion have been found in terms of the integrals of the mode functions and closed form terms. These matching coefficents have been used to numerically compute part of the {\it in} mode function on the future horizon of the black hole.

The second part of the method is the renormalization of the stress-energy tensor which involves taking the difference between the stress-energy tensor for the "in" state in the collapsing null shell spacetime and that for Unruh state in the Schwarzschild spacetime. Finally, we reviewed the computation of the stress-energy tensor in the corresponding 2D case using the aformentioned method.
\section{Aknowledgement}
P. R. A. would like to thank Eric Carlson, Charles Evans, Adam Levi, and Amos Ori for helpful conversations and Adam Levi for sharing some of his numerical data. A.F. acknowledges partial financial support by the Spanish grants FIS2017-84440-C2-1-P funded by MCIN/AEI/10.13039/501100011033 ''ERDF A way of making Europe'', Grant PID2020-116567GB-C21 funded by  MCIN/AEI/10.13039/501100011033, and the project PROMETEO/2020/079 (Generalitat Valenciana). This work was supported in part by the National Science Foundation under Grants No.  PHY-1308325, PHY-1505875, and PHY-1912584 to Wake Forest University. Some of the numerical work was done using the WFU DEAC cluster; we thank the WFU Provost's Office and Information Systems Department for their generous support.
\bibliographystyle{ws-procs961x669} 
\bibliography{ws-pro-sample}

\begin{thebibliography}{0}
\bibitem{fawcett} M. S. Fawcett, Commun.\ Math.\ Phys {\bf 89} {\rm}, 103 (1983).
\bibitem{howard-candelas} K.~W.~Howard and P. Candelas, Phys. Rev. Lett. {\bf 53}, 403 (1984).
\bibitem{howard} K.~W.~Howard, Phys. Rev. D {\bf{30}}, 2532 (1984).
\bibitem{jensen-ottewill} B. P. Jensen and A. Ottewill, Phys. Rev. D {\bf 39}, {\rm 1130} (1989).
\bibitem{jmo} B. P. Jensen, J. G. McLaughlin, and A. C. Ottewill,
 Phys. Rev. D {\bf 43}, 4142 (1991).
\bibitem{jensen-et-al} B. P. Jensen, J. G. Mc Laughlin, and A. C. Ottewill,  Phys. Rev. D {\bf 45}, 3002 (1992).
\bibitem{ahs1} P. R. Anderson, W. A. Hiscock, and D. A. Samuel, Phys. Rev. Lett. {\bf 70}, 1739 (1993).
\bibitem{ahs2}  P. R. Anderson, W. A. Hiscock, and D. A. Samuel, Phys. Rev. D {\bf 51}, 4337 (1995).
\bibitem{ahl} P. R. Anderson, W. A. Hiscock, and D.~J.~Loranz, Phys. Rev. Lett. {\bf 74}, 4365 (1995).
\bibitem{choag} E. D. Carlson, W. H. Hirsch, B. Obermayer, P. R. Anderson, and P. B. Groves, Phys. Rev. Lett. {\bf 91}, 051301 (2003).
\bibitem{abf} P. R. Anderson, R. Balbinot, and A. Fabbri, Phys. Rev. Lett. {\bf 94}, 061301 (2005).
\bibitem{breen-ottewill} C. Breen and A. C. Ottewill, Phys. Rev. D {\bf 85}, 084029 (2012).
\bibitem{levi-ori} A. Levi and A. Ori, Phys. Rev. Lett. {\bf 117}, 231101 (2016).
\bibitem{levi} A. Levi, Phys. Rev. D {\bf 95}, 025007 (2017).
\bibitem{Zilberman-Levi-Ori} N. Zilberman, A. Levi, A. Ori, Phys. Rev. Lett.  {\bf 124}, 171302 (2020).
\bibitem{duffy-ottewill} G. Duffy and A. C. Ottewill, Phys. Rev. D {\bf 77}, 024007 (2008).
\bibitem{levi-et-al-kerr} A. Levi, E. Eilon, A. Ori, and M. van de Meent, Phys. Rev. Lett. {\bf 118}, 141102 (2017).
\bibitem{ourpaper} P. R. Anderson, S Gholizadeh Siahmazgi, R. D. Clark, and A.Fabbri,  Phys.\ Rev.\ D {\bf 102}, 125035 (2020).
\bibitem{Fabbri:2005mw}
  A.~Fabbri and J.~Navarro-Salas, {\it Modeling black hole evaporation} (Imperial College Press, London, UK, 2005).
\bibitem{m-p} S.~Massar and R.~Parentani,   Phys.\ Rev.\ D {\bf 54}, 7444 (996).
\bibitem{rigorous} P. R. Anderson, A. Fabbri, and R. Balbinot,  Phys.\ Rev.\ D {\bf 91}, 064061 (2015).
\bibitem{christensen-76} S. M. Christensen,  Phys.\ Rev.\ D {\bf 14}, 2490 (1976).
\bibitem{mirror-bh} M. R.R. Good, P. R. Anderson, and C. R. Evans,   Phys.\ Rev.\ D {\bf 94}, 065010 (2016).
\bibitem{late-time} P. R. Anderson, R. D. Clark, A. Fabbri, and M. R. R. Good, Phys.\ Rev.\ D {\bf 100}, 061703(R) (2019).
\bibitem{hiscock} W. A. Hiscock,  Phys. Rev. D {\bf 23}, 2813 (1981).
\bibitem{Hawking:1974sw}
S. W. Hawking,
{\href{https://link.springer.com/article/10.1007/BF02345020}{Commun. Math. Phys. \textbf{43} 199 (1975)}}.
\bibitem{Unruh-1976} W. G. Unruh, Phys. Rev. D {\bf 14}, 870 (1976).
\bibitem{elster} T. Elster, Phys. Lett. {\bf 94A}, 205 (1983).
\bibitem{akhmedov-godazgar-popov} E. T. Akhmedov, H. Godazgar, and F. K. Popov, Phys. Rev. D {\bf 93}, 024029 (2016).
\bibitem{Davies-Fulling-Unruh} P. C. W. Davies, S. A. Fulling, and W. G. Unruh,
 Phys. Rev. D {\bf 13}, 2720 (1976).
\bibitem{BEC-2013} P. R. Anderson, R. Balbinot, A. Fabbri, and R. Parentani,  Phys.\ Rev.\ D {\bf 87}, 124018 (2013).
\end{thebibliography}

\end{document}